\documentclass[reprint,amsmath,amssymb,aps,prl,floatfix,twocolumn,superscriptaddress]{revtex4-2}

\usepackage[dvipsnames]{xcolor}

\usepackage{hyperref}
\hypersetup{linktocpage,
	pdftoolbar=true,        
	pdfmenubar=true,        
	pdffitwindow=false,     
	pdfstartview={FitH},    
	pdftitle={title},    
	pdfauthor={Gerard Valentí-Rojas},     
	pdfsubject={Subject},   
	pdfcreator={Gerard Valentí-Rojas},   
	pdfproducer={Gerard Valentí-Rojas}, 
	pdfnewwindow=true,      
	colorlinks=true,       
	linkcolor={red!50!black},          
	citecolor={blue!50!black},        
	filecolor={blue!80!black},      
	urlcolor={blue!80!black}           
}

\usepackage{csquotes}
\usepackage[inline]{enumitem}
\usepackage{graphicx}
\usepackage{ifthen}
\usepackage{amsmath}
\usepackage{amsthm}
\usepackage{pifont}
\usepackage{array}
\usepackage{siunitx}
\usepackage{xstring}
\usepackage{multirow}
\usepackage{datetime}
\usepackage{etoolbox}

\usepackage{slashed}

\usepackage{bm}
\usepackage{braket}

\usepackage{booktabs}

\usepackage{cleveref}

\usepackage{framed}

\definecolor{shadecolor}{rgb}{0.95,0.95,0.95}

\usepackage{newtxtext,newtxmath}

\usepackage{graphicx}
\usepackage{dcolumn}
\usepackage{url}

\begin{document}
\title{\textcolor{red!50!black}{Dual approach to soft-core anyonic Lieb-Liniger fluids}}
\author{Gerard Valent\'i-Rojas}
\email[Correspondence: ]{gerard.valenti-i-rojas@institutoptique.fr}
\affiliation{Naquidis Center, Institut d'Optique Graduate School, 91127, Palaiseau, France}
\affiliation{SUPA, Institute of Photonics and Quantum Sciences, Heriot-Watt University, Edinburgh, EH14 4AS, United Kingdom}

\author{Patrik \"Ohberg}
\affiliation{SUPA, Institute of Photonics and Quantum Sciences, Heriot-Watt University, Edinburgh, EH14 4AS, United Kingdom}

\date{\today}
\begin{abstract}
The identity of quantum matter can be effectively altered by means of gauge fields. In two spatial dimensions this is illustrated by the Chern-Simons flux-attachment mechanism, but such a mechanism is not possible in lower dimensions. Here, we study a one-dimensional interacting Bose gas in the presence of a gauge field. This model can be explicitly mapped into an interacting anyonic system by a large gauge transformation, indicating a statistical transmutation analogous to that of Chern-Simons. The Bogoliubov spectrum in the weakly-interacting limit reveals the presence of a roton minimum arising from the statistical interaction. At a mean-field level chiral solitons are recovered. Should these be understood as quantum bound states, it is natural to interpret them as corresponding to localised anyonic quasiparticles. Hydrodynamic arguments highlight the presence of dispersive chiral shock waves in the propagation of a wavepacket due to a Riemann-Hopf nonlinearity. Numerical calculations show the presence of both chiral soliton trains and shock waves.
\end{abstract}
\maketitle

\textit{\textbf{Introduction.}--- }
 Quantum particles have an intrinsic identity. Their properties can be summarised in a handful of magnitudes accounting for the most relevant degrees of freedom available. In a condensed matter context, these are mass ($m$), charge ($q$), spin ($s$), statistics ($\eta$) and lifetime ($\tau$), among few others. Out of these properties, spin and statistics encode the \textit{quantum identity} of matter as they have no classical analogue. Quasiparticles appear as localised excitations in interacting many-body systems and possess similar qualities $(m^{*},q^{*},s^{*},\eta^{*},\tau^{*})$ with the important difference that they are emergent, as opposed to the prior, which are fundamental. 

 For a long time, quasiparticles have been found to inherit part of their identity from the underlying fields due to adiabatic continuity \cite{anderson2018basic}. Anyons \cite{leinaas1977theory,goldin1981representations,wilckez1982flux,wilczek1982fraction}, however, constitute an important counterpoint to this framework, as quasiparticles with widely different properties from the original microscopic interacting system. The exotic nature of anyons, as compared to conventional quantum particles, has opened entirely new research avenues and already constitutes a key element for near-future technologies \cite{kitaev2003fault,pachos2012introduction,nayak2008non,terhal2015error}.

While systems giving rise to anyonic and fractionalised excitations have been thoroughly explored \cite{wilczek1990book,fradkin2023field,moessner2021topological,sachdev2023quantum}, a \textit{general} physical understanding for how conventional quantum systems can yield exotic matter featuring anyons is lacking. One such mechanism in two spatial dimensions is flux attachment \cite{wilckez1982flux,wilczek1982fraction}, a by-product of a many-body Aharonov-Bohm effect in which electrically-charged particles capture magnetic flux quanta and become Chern-Simons gauge-dressed entities with transmuted statistics \cite{polyakov1988fermi}. In one spatial dimension anyonic statistics is still allowed, but the mechanism of flux attachment is no longer possible as there is no notion of magnetic flux. The physical intuition for anyons in this context is much less clear. This is manifest in the lack of consensus for what an anyon \textit{actually is} in setups such as a one-dimensional quantum fluid \footnote{An exception to this is the case of Majorana zero modes \cite{sarma2015majorana,lutchyn2018majorana}, for which much more consensus exists.}. More concretely, given an initial one-dimensional continuous system whose components have a given set of conventional $(m,q,s,\eta,\tau)$, how does one obtain a fluid with anyonic quasiparticles and what do they look like? Is there a notion of a statistical gauge field in this context in analogy to well-known Fractional Quantum Hall (FQH) fluids \cite{jain2007composite,fradkin2013field,ezawa2013quantum}?

Here, we explore an explicit example of a ``conventional'' quantum system with $(\eta)$ that becomes ``exotic'' with $(\eta^{*})$  by coupling it to a certain gauge field. More concretely, we discuss the connection between a bosonic fluid in one spatial dimension and its composite dual model, the soft-core anyonic Lieb-Liniger model \cite{kundy99anyons,pactu2007correlation,Batchelor2008}. We comment on phenomenological features of the model interpreted in the bare (bosonic) basis. This serves as a concrete application of the composite particle duality \cite{valenti2023composite}, which argues that the notion of statistical gauge fields is valid in any dimension. We discuss the formation of exotic solitons interpreted as bound states, shock waves at long times, and asymmetric dynamics.\\

\textit{\textbf{Dual Quantum Fluids.}--- }
Our starting point is a locally-interacting bosonic fluid in one spatial dimension. An implementation and manipulation of such a model can be achieved, for instance, with ultracold gases in an optical waveguide \cite{atom2022tronics}. According to the composite particle duality \cite{valenti2023composite}, the coupling of such a conventional state of quantum matter to an appropriate statistical gauge field allows the system to be exactly transformed into an anyonic system. More concretely, the bosonic quantum fluid with Lagrangian density
\begin{equation}\label{eq:boson}
    \mathcal{L}_{\text{B}}= i\hbar \,\hat{\Psi}^{\dagger}D_{t}\hat{\Psi} - \frac{\hbar^{2}}{2m} \big(D_{x}\hat{\Psi}\big)^{\dagger} \big(D_{x}\hat{\Psi}\big) - \frac{g}{2}\, \hat{\Psi}^{\dagger}\hat{\Psi}^{\dagger}\hat{\Psi}\hat{\Psi} 
\end{equation}
and conventional equal-time commutation relations
\begin{flalign}
	&\big[\hat{\Psi}(x)\,,\hat{\Psi}(x') \big] =\big[\hat{\Psi}^{\dagger}(x)\,,\hat{\Psi}^{\dagger}(x') \big]= 0 \;,\\
	&\big[\hat{\Psi}\,(x)\, ,\hat{\Psi}^{\dagger}(x') \big]= \delta\,(x-x')\;,
\end{flalign}
is equivalent to a soft-core anyonic Lieb-Liniger gas
\begin{equation}
\tilde{\mathcal{L}}_{\text{C}} = i\hbar\, \hat{\Psi}_{\text{C}}^{\dagger}\,\partial_{t} \hat{\Psi}_{\text{C}}  - \frac{\hbar^{2}}{2m} \, \partial_{x} \hat{\Psi}_{\text{C}}^{\dagger} \,\partial_{x} \hat{\Psi}_{\text{C}} -\frac{g}{2} \,\hat{\Psi}_{\text{C}}^{\dagger}\hat{\Psi}_{\text{C}}^{\dagger} \hat{\Psi}_{\text{C}} \hat{\Psi}_{\text{C}}\; ,
\end{equation}
with the composite field obeying the anyonic algebra
\begin{flalign}
	&\hat{\Psi}_{\text{C}}^{\dagger}(x) \,\hat{\Psi}_{\text{C}}^{\dagger}(x') - e^{\,\frac{i}{\hbar}\gamma \,\text{sgn}\,(x-x')} \,\hat{\Psi}_{\text{C}}^{\dagger}(x')\, \hat{\Psi}_{\text{C}}^{\dagger}(x) = 0\,, \\
	&\hat{\Psi}_{\text{C}} (x) \,\hat{\Psi}_{\text{C}}^{\dagger}(x') - e^{\,-\frac{i}{\hbar}\gamma \,\text{sgn}\,(x-x')} \,\hat{\Psi}_{\text{C}}^{\dagger}(x')\, \hat{\Psi}_{\text{C}} (x) = \delta\,(x-x')\,,
\end{flalign}
upon redefining the bosonic field as a composite via the large gauge transformation
\begin{equation}
   	\hat{\Psi}\,(t,x) = e^{\,\frac{i}{\hbar}\hat{\Phi}(t,x)}\, \hat{\Psi}_{\text{C}}\,(t,x) \;,
\end{equation}
where $\hat{\Phi}$ is a disorder field or a generalised Jordan-Wigner string
\begin{equation}
	\hat{\Phi}\,(t,x) = \gamma \int_{-\infty}^{\infty} dx'\;\Theta\,(x- x')\,\hat{n}\,(t,x') \;,
\end{equation}
and where the Heaviside step function $\Theta (x)$ plays the role of a kink. The bosonic matter field in Eq. \eqref{eq:boson} is minimally coupled to a statistical gauge field $\hat{a}_{\mu}$ via the gauge-covariant derivative $D_{\mu} = \partial_{\mu} -i\hbar^{-1} \hat{a}_{\mu}\,$, where the form of the gauge field is set by the conserved current constraint
\begin{equation}
    \hat{J}^{\,\mu} = \gamma^{-1}\epsilon^{\,\mu\nu}\hat{a}_{\nu}\;\;\;\;\;\;\;\text{and} \;\;\;\;\;\;\; \partial_{\mu}\,\hat{J}^{\,\mu} = 0\;,
\end{equation}
and where the covariant Noether current density is defined in the conventional way
\begin{equation}
\hat{J}^{\,\mu}\,(t,x) = \bigg(\,\hat{\Psi}^{\dagger} \hat{\Psi}\,, \;\frac{\hbar}{2mi}\, \Big[\hat{\Psi}^{\dagger}D_{x}\,\hat{\Psi} - (D_{x}\,\hat{\Psi})^{\dagger}\,\hat{\Psi}\Big] \,\,\bigg)\;.
\end{equation}

Since both systems are equivalent, it is a matter of taste whether to work with the original or transformed formulation. We refer to this as working on either the bare ($\text{B}$) or the composite ($\text{C}$) basis. While the composite basis might prove useful for theoretical calculations \cite{kundy99anyons,pactu2007correlation,Batchelor2008}, ultimately the eventual experimental system is bosonic, so the latter basis is preferable to understand the system from an implementation standpoint.\\

\begin{figure}[h]
    \includegraphics[width=0.47\textwidth]{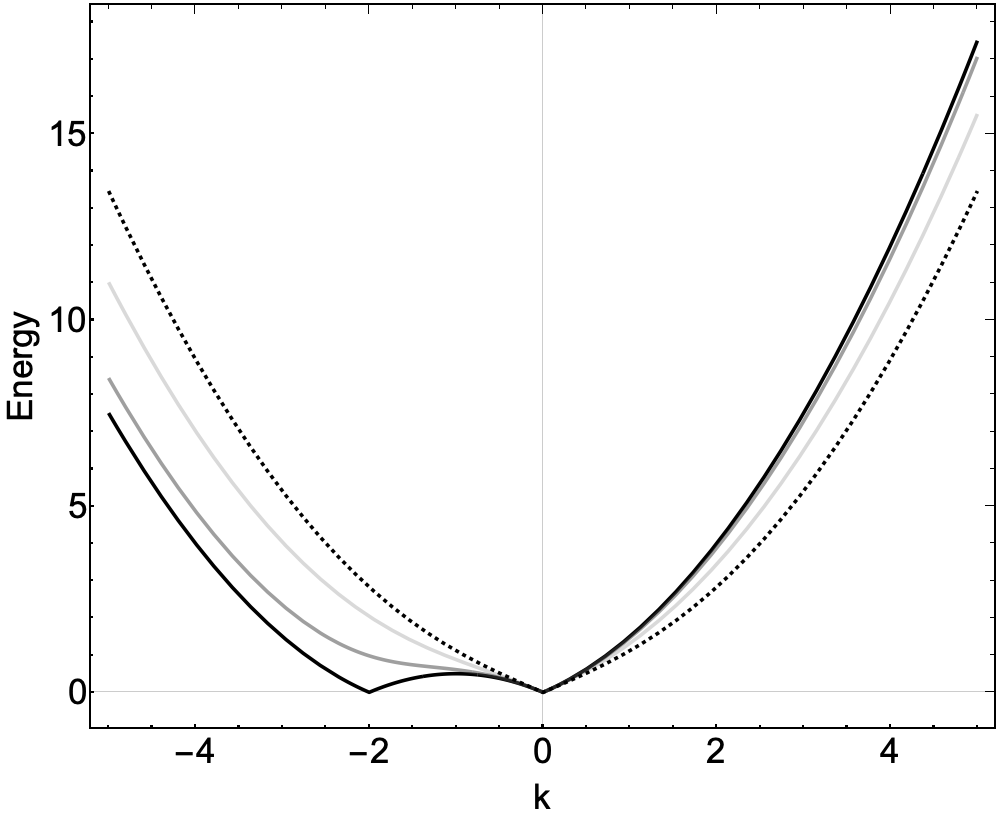}
    \caption{Bogoliubov spectrum in dimensionless units for $\gamma=0$ (dashed) and for increasing values of $0< \gamma \leq 1/4$ (gray). For non-vanishing values of the statistical interaction, an asymmetry in the spectrum develops and a roton minimum appears. As $\gamma/g_{\text{eff}}$ is increased, the roton is softened until it reaches zero energy. For $\gamma > 1/4$, the spectrum becomes complex-valued (not shown).}
   \label{fig:rotonbogo}
\end{figure}

\textit{\textbf{Elementary Excitations.}--- }
The Hamiltonian formulation in the bare basis can be written as $H = H_{\text{LL}} + H_{\text{G}}\,$, where
\begin{equation}
H_{\text{LL}} = \int dx\; \bigg(\frac{\hbar^{2}}{2m} \partial_{x}\hat{\Psi}^{\dagger} \partial_{x}\hat{\Psi} +\frac{g}{2} \hat{\Psi}^{\dagger}\hat{\Psi}^{\dagger} \hat{\Psi} \hat{\Psi} \bigg)
\end{equation}
refers to the usual Lieb-Liniger model, and
\begin{equation}
H_{\text{G}} = H_{\chi} + H_{\text{3b}} = 2\gamma\int dx\; \bigg(  \colon \!  \hat{n}\,\hat{j}_{x} \colon \! -\frac{3\gamma}{4m}\hat{\Psi}^{\dagger}\hat{n}^{2}\hat{\Psi} \bigg) 
\end{equation}
results from the coupling to the statistical gauge field. Here, $(\colon \!  \bullet \colon \!)$ denotes bosonic normal ordering and $\hat{j}_{x}$ is the ungauged current density. We consider the weakly-repulsive Bogoliubov regime and calculate the excitation spectrum in the conventional way \cite{cazalilla2011review,risti2016decay}. The key novel term is the density-current chiral interaction
\begin{flalign}
    H_{\chi} = \frac{2\gamma}{\text{Vol}} \sum_{k_{1},k_{2},q} \bigg( \frac{\hbar k_{1} }{m} \bigg) \;\hat{b}^{\dagger}_{k_{2}+q} \hat{b}^{\dagger}_{k_{1} - \frac{q}{2}} \hat{b}_{k_{1}+\frac{q}{2}} \hat{b}_{k_{2}} \;,
\end{flalign}
which explicitly breaks \textit{parity} ($\mathcal{P}$) and \textit{time-reversal} ($\mathcal{T}$) symmetries. The other local interaction terms in $ H_{\text{3b}}$, namely $\sim \gamma^{2} \colon \hat{n}^{2}\colon $ and $\sim \gamma^{2}\colon  \hat{n}^{3} \colon \,$ merely renormalise the conventional contact interparticle interactions in the Bogoliubov limit to $g \rightarrow  g_{\text{eff}}\,$. Like in conventional Bogoliubov theory for the dilute Bose gas, we expand in low momenta and consider only zero total momentum state $\Delta k = k_{\text{in}} - k_{\text{out}}= 0$. We find that $H_{\chi}$ merely modifies the standard contact interaction by
\begin{equation}
    g \longrightarrow \tilde{g}(k) = g_{\text{eff}}\,\bigg(1+\gamma\frac{4\hbar k}{m g_{\text{eff}}}\bigg)\;.
\end{equation}
Hence the Bogoliubov spectrum reads as
\begin{equation}
    \mathcal{E}_{k} = \sqrt{\bigg( \frac{\hbar^{2} k^{2}}{2m} \bigg)^{2}+ \tilde{g}(k)\,\frac{n_{0} \,\hbar^{2} k^{2}}{m}\,\;\;} \;\;,
\end{equation}
where $n_{0}= N/\text{Vol}$ is the number density. This dispersion relation can be seen in Figure \ref{fig:rotonbogo}. The spectrum has a characteristic phonon-maxon-roton shape, unlike the usual weakly-interacting Bose gas.
We appreciate three characteristic features:
 \begin{itemize}
    \item There are gapless phonons.
    \item There is a roton minimum for $\text{sgn}(\gamma k)<0$.
    \item It is asymmetric in $k$.
\end{itemize}   

\paragraph{On phonons.} Some insight can be gained from comparing this spectrum with that of the Zhang-Hansson-Kivelson (ZHK) model \cite{girvin1987off, zhang1989effective,zhang1995chern,read1989order} for a charged superfluid coupled to a Chern-Simons gauge field in two spatial dimensions. We observe a qualitative difference with respect to phonons. Here, phonons do not acquire a topological mass gap, so there is no Anderson-Higgs mechanism. In one dimension there is no sense of a magnetic field, so no notion of cyclotron motion. Thus, the cyclotron frequency, that determines the size of the magnetophonon gap in two spatial dimensions \cite{girvin1986magneto,zhang1995chern}, vanishes in the one-dimensional limit. This implies that the system is compressible, unlike a FQH fluid.\\

\paragraph{On asymmetry.} The spectrum shows some similarities with that of a spin-orbit-coupled condensate \cite{khamechi2014measure,chin2015roton}. Given the spectral asymmetry, two critical velocities corresponding to $\pm k$ flow shall be defined with respect to the ($+$) phonon  and ($-$) roton. Landau's criterion for superfluidity has to be modified accordingly due to the breaking of parity. This is a recurrent feature in spin-orbit-coupled condensates \cite{chin2015roton,zhu2012exotic}. The speed of sound is also affected as a consequence. The particular case of a superfluid coupled to a  certain class of density-dependent gauge fields has been studied at a classical hydrodynamic level in Ref. \cite{buggy2020hydrodynamical}. Buggy finds an extended analytical expression for the speed of sound in the condensate that is dependent on the statistical parameter. The propagation of phonons in such systems is, in general, not only asymmetric but also highly non-trival.\\ 

\paragraph{On rotons.} As the ratio $\gamma/g_{\text{eff}}$ becomes of order one, a roton minimum develops. This feature is also found in the ZHK model. As the statistical interaction increases the roton is softened down to $\mathcal{E}_{k} = 0$ at finite $k<0$. For sufficiently large values of the ratio, the spectrum becomes complex-valued, signalling an instability. The roton satisfies the usual dispersion
\begin{equation}\label{eq:roton}
    \mathcal{E}_{\text{rot}}(k) = \frac{\hbar^{2}(k -k_{r})^{2}}{2m^{*}} + \Delta
\end{equation}
with gap $\Delta$ and effective mass $m^{*}$. At $k=k_{r}$ the roton has zero net group velocity at non-vanishing momentum. The gap of the roton minimum depends on the relative strength of the statistical and interparticle interactions. In this sense, the statistical gauge field yields an effect similar to conventional long-range interactions, such as in dipolar gases \cite{giovanazzi2004instabilities,chomaz2018observation}. We find that for the ratio $\gamma/g_{\text{eff}} \rightarrow 1/4$ the gap vanishes $\Delta \rightarrow 0$, signalling a practically null energy cost for the creation of this collective mode at finite wave-number.  \\ 

\paragraph{On solitons.} Solving the classical equations of motion for the matter field in terms of a generalised chiral Gross-Pitaevskii equation, reveals unusual soliton solutions \cite{aglietti1996anyons,nishino1998chiral,griguolo1998chiral,jackiw1997review,edmonds2013simulating,chisholm2022encoding,min1996scattering}, which are stable for one chirality, but unstable for the other. A bright chiral soliton has been experimentally implemented and observed by Frölian \textit{et al. }\cite{frolian2022realising}. The chiral aspect of the soliton appears as an  instability to backscattering, something which both experiments \cite{frolian2022realising} and numerical simulations in the current work confirm. Such a solution should be robust upon additional $\mathcal{N}-$body interactions \cite{harikumar1998chiral,kivshar1989dynamics}. This soliton is the one-dimensional analogue of an anyonic ZHK vortex. It is plausible to conclude that it is a gauge-charge composite, namely a linear anyon.\\

\begin{figure}[h]
    \includegraphics[width=0.49\textwidth]{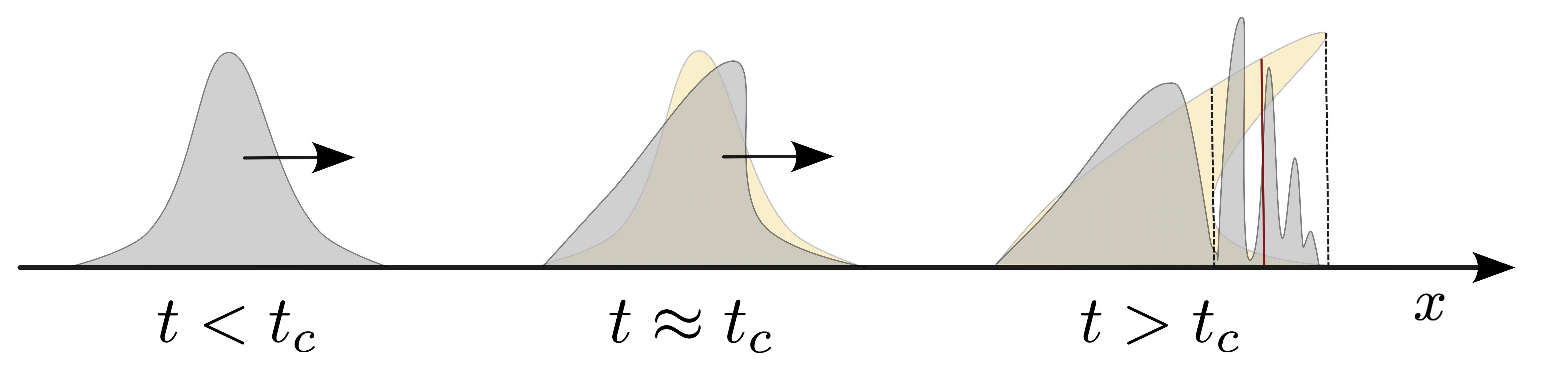}
    \caption{Schematic of a dispersive shock wave formation by a Riemann-Hopf nonlinearity. A wavepacket propagates to the right, with initial stages of the dynamics being approximately linear. Due to a Riemann-Hopf nonlinearity, a density overhang is created in the direction of motion. At breaking time $t=t_{c}$, the Riemann-Hopf term creates a mathematical gradient singularity known as a \textit{gradient catastrophe}. This manifests in the formation of a shock front. The nonlinearity is regularised by higher order dispersive terms creating highly oscillatory behaviour which can yield solitons in its wake.}
   \label{fig:gradient}
\end{figure}
\begin{figure*}[th]
\centering
    \includegraphics[width=0.47\textwidth]{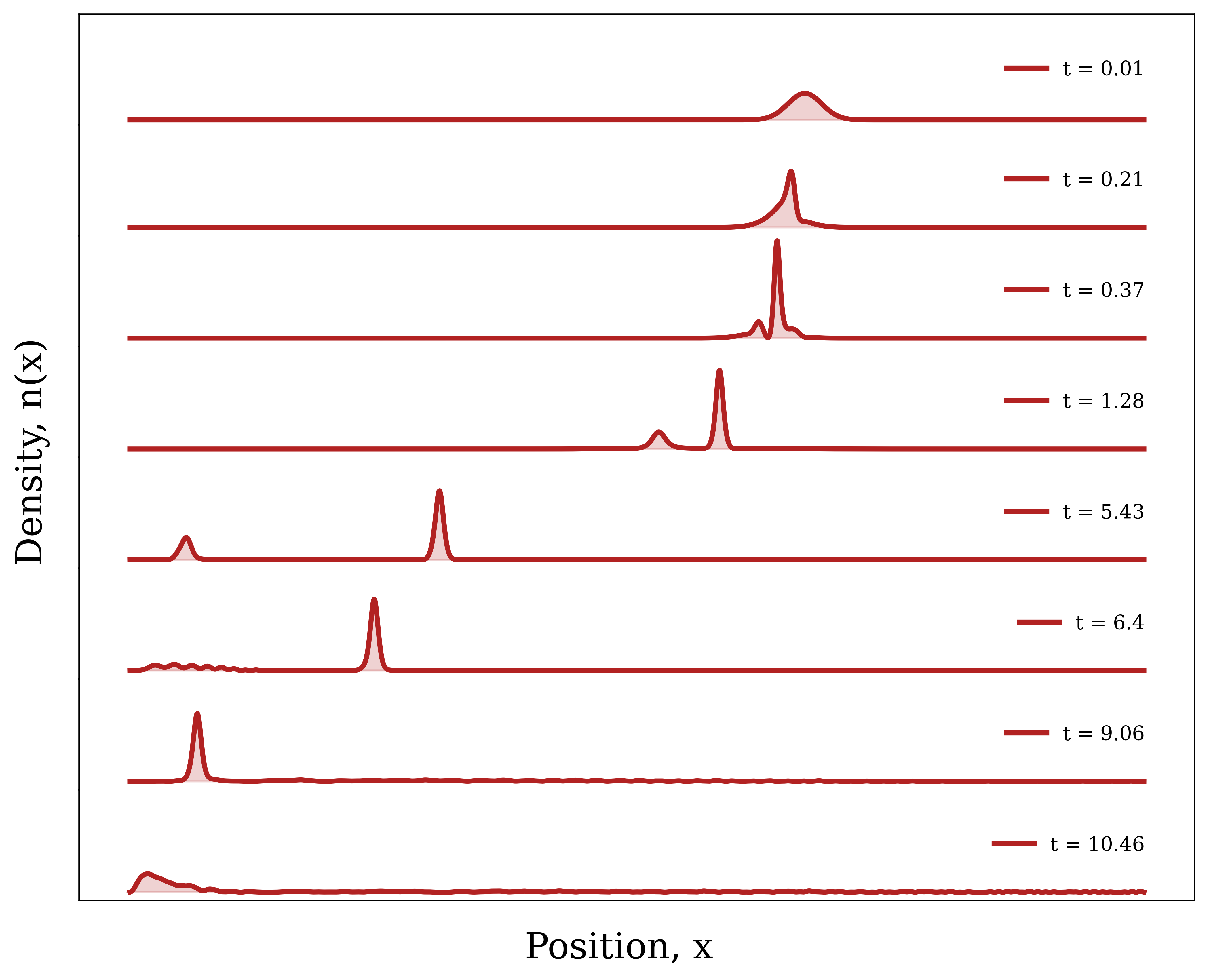}
    \hspace{0.45cm}
    \includegraphics[width=0.47\textwidth]{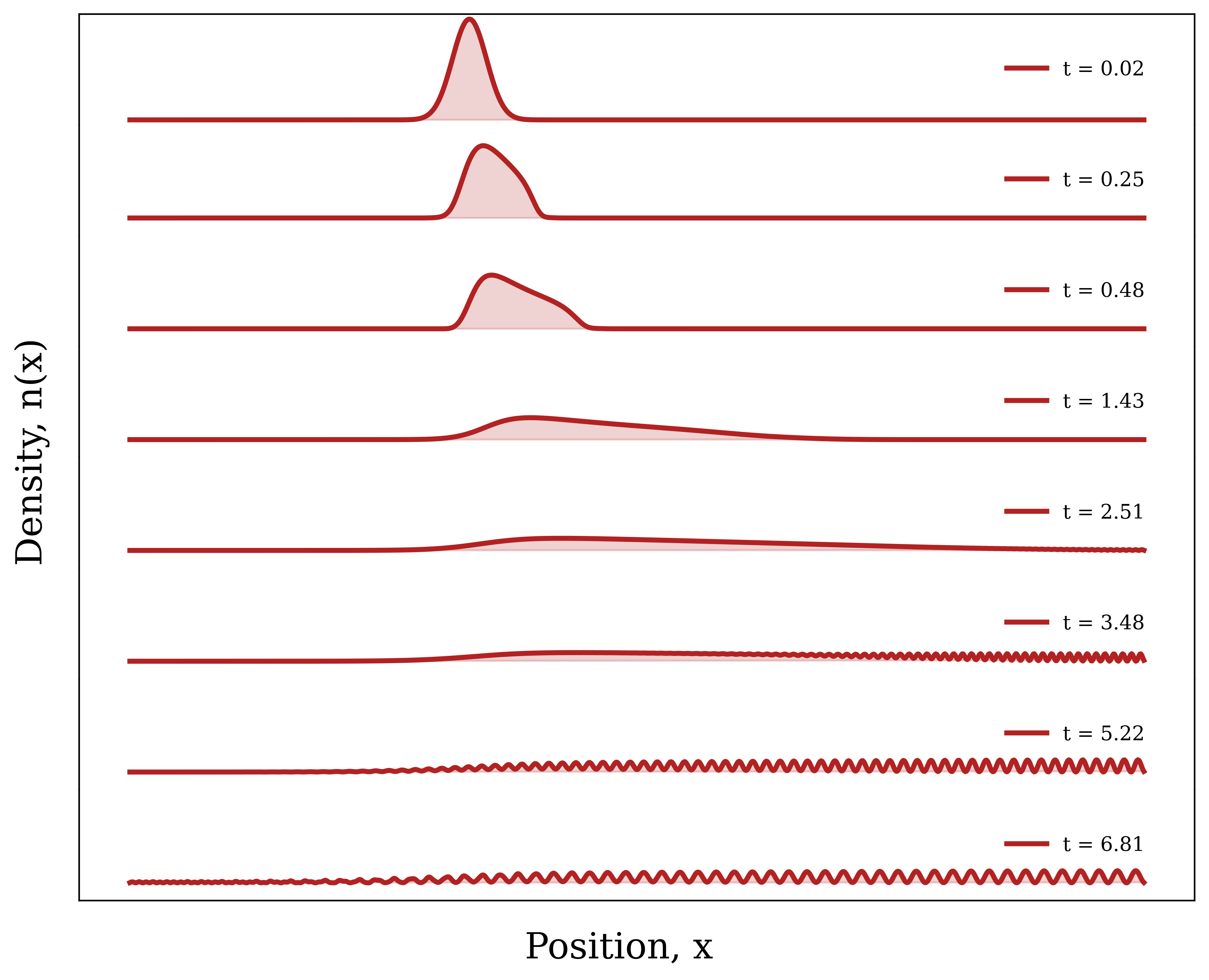}
    \caption{Anomalous reflection dynamics for density-current nonlinear wavepacket boosted with a $\lvert k_{0}\rvert = 3$ momentum kick and $\gamma = 2$. (\textit{Left}) Creation of a $k_{0}<0$ chiral soliton train with $N=2$ solitons. The ``small'' soliton propagates at higher group velocity than the ``big'' one. When in contact with the wall and backscattered, the solutions are unstable. For times $t\sim 7-8$, the backscattered wavepacket and the ``big'' soliton interact. The soliton solution is robust against this perturbation. At late times, the ``big'' soliton also bounces off the wall and gets destroyed. (\textit{Right}) For $k_{0}>0$ we observe an asymmetric expansion followed by shock wave after reflection on a wall. The highly oscillatory behaviour is characteristic of dispersive contributions regularising a gradient catastrophe. The shock wave is robust at late times and results in a regular density modulation. All parameters are in dimensionless units.}
   \label{fig:same}
\end{figure*}

\paragraph{On Shock Waves.} Regarding the dynamics of the system, even the simplest semi-classical hydrodynamic analysis reveals that the current nonlinearity will give rise to shock waves. In other words, for moderate values of the parameter $\gamma$ a propagating wavepacket will eventually develop a density overhang and experience a wave-breaking effect of a form similar to that described in Figure \ref{fig:gradient}. The reason being the presence of a dominant Riemann-Hopf nonlinearity in addition to higher order terms in the conservation of current
\begin{equation}\label{eq:hydrogauge}
 0 = \partial_{t} n + \partial_{x}\,\big(n\,\tilde{v}_{s} \big) =  \partial_{t} n + \partial_{x}\,\big(n\,v_{s} \big) - \frac{2\gamma}{m}\,n \,\partial_{x} n \;,
\end{equation}
where the last term creates a Riemann wave. The gauged superfluid velocity is defined in the conventional way
\begin{equation}
   \tilde{v}_{s} = \frac{\hbar}{m}\,\bigg(\partial_{x}\theta-\frac{1}{\hbar} a_{x} \bigg) = v_{s}-\frac{\gamma}{m} n\;.
\end{equation}
The \textit{gradient catastrophe} will be driven by the statistical parameter $\gamma$, which governs the strength of the nonlinearity. The catastrophe will be regularised by higher order dominant terms in the corresponding Euler equation. We expect this regularisation to be dispersive \cite{el2016dispersive} and the shock wave to be chiral. The existence of shock waves at long times is generic, while its characterisation requires detailed numerical and analytical machinery beyond the scope of this work. However, preliminary confirmation of this intuition can be gained by generating a momentum kick $k_{0}$ in the wavepacket near a wall \cite{dubessi2021shock}. When the kick and the gauge coupling have the same sign, i.e. $\text{sgn}(\gamma k_{0})>0$, we observe an asymmetric expansion and the creation of a robust density wave modulation after the contact with a wall (see Figure \ref{fig:same} right). For the case of opposite signs, we observe the creation of a chiral soliton train (see Figure \ref{fig:same} left). For the latter case, solitons get destroyed one by one upon backscattering with the wall.

We would also like to draw attention to the fact that the last term in Eq. \eqref{eq:hydrogauge} can be seen as a source term to the conventional current conservation. Provided that this contribution is generated by a gauge field, this source term can be interpreted as an unusual chiral anomaly in a non-Galilean fluid.\\

\textit{\textbf{Discussion and Conclusions.}--- } 
A paradigmatic example of conventional quantum matter, such as an interacting Bose gas can be exactly mapped into an anyonic quantum fluid upon coupling to a certain gauge field. This situation is well-understood in two spatial dimensions for the case of FQH fluids, where the Chern-Simons mechanism of flux attachment transmutes the quantum identity of the many-body system. This might even be experimentally tunable \cite{valenti20synthetic}.

We have studied a spatially one-dimensional interacting bosonic system. Strictly speaking, there is no Chern-Simons theory in one spatial dimension, no mechanism of flux attachment, no cyclotron motion, and no sense of topological quantisation. However, we advocate for the existence of a remnant, a gauge field that plays an analogous role in altering the commutation relations of the underlying matter fields, i.e. statistically transmuting their quantum identity. This lower dimensional analogue can be understood as arising from a \textit{chiral axion} term \cite{valenti2023composite}, which is also the natural dimensional reduction of the Abelian Chern-Simons term.

Because of the low dimensionality of the system, the gauge field can be fully eliminated leading to a residual statistical interaction in the form of a local $\mathcal{P}$ and $\mathcal{T}$ breaking interparticle interaction term. Its effect is significant, as it gives rise to important elementary excitations not present in the usual non-gauged bosonic model, namely chiral solitons and rotons. These are reminiscent of the Chern-Simons magnetorotons and anyonic ZHK vortices. An immediate natural question is whether these chiral solitons have fractional charge and statistics. We argue that such classical solutions can be understood as quantum bound states, and correspond to a gauge-matter composite that can be identified with one-dimensional anyons. Of course, a more convincing answer that addresses this issue should come from a fully quantum treatment of these solutions, which we did not pursue in this work. Other important aspects of the current model are \textit{(i)} the absence of an Anderson-Higgs mechanism, and thus, the presence of massless Goldstone modes; and \textit{(ii)} the creation of chiral dispersive shock waves, and the possibility of generating chiral solitons trains or chiral density waves in the propagation of the fluid at moderate values of the gauge coupling. All of the above is within experimental reach with current state-of-the-art techniques in ultracold atoms. \\

\textit{\textbf{Further Considerations.}--- } It is understood from exact results in the conventional Lieb-Liniger model \cite{lieb1963i,lieb1963ii,lang2017ground} that exact results beyond a Bogoliubov analysis reveal the existence of a ``replica'' excitation branch in addition to the one already discussed. These two branches are in one-to-one correspondence with fermionic particle-hole branches for excitations. In other words, a more profound analysis reveals particle-like (Lieb I mode) and hole-like (Lieb II mode) excitations. The Lieb I mode corresponds to the usual Bogoliubov phononic branch, while the Lieb II mode can be identified with the mean-field Gross-Pitevskii dark soliton branch \cite{ishikawa1980solitons,kopycinski2022beyond}. The exact Bose-Fermi correspondence \cite{granet2022duality} between the Lieb-Liniger and Cheon-Shinegara fluids \cite{cheon1999fbduality} might be the way out as it provides a duality between two models with manifestly different quantum identities.

The exact correspondence between the statistically-gauged interacting Bose gas and the soft-core anyonic Lieb-Liniger fluid extends the above notions. Guided by intuition, we are tempted to conjecture a soliton-particle excitation correspondence between bare-composite bases. This is suggestive of an order-disorder duality in which one can either choose to see a system of bosons with solitonic excitations or, treating those solitons as composite entities \footnote{We note here that the idea of interpreting certain chiral solitons as composite particles has been discussed in other contexts \cite{alkofer1996baryons}. More generally, this goes in the same spirit as the sine-Gordon/massive-Thirring correspondence.}, reinterpret the theory as a gas of interacting effective anyonic particles. Alas, more detailed analysis should be conducted.\\ 

\textit{\textbf{Acknowledgments.}--- } 
We warmly thank A. Celi and L. Tarruell for useful discussions. G.V-R. acknowledges financial support from EPSRC CM-CDT Grant No. EP/L015110/1. 


\bibliography{references}
\bibliographystyle{apsrev4-2}

\end{document}